# The Astri Mini-Array of Cherenkov Telescopes

**Andrea Giuliani**[a,][*] **for the ASTRI Project**[b]

[a]*INAF – IASF Milano,*
*Via Alfonso Corti 12, I-20133 Milano, Italy*

[b]*http://www.astri.inaf.it*

*E-mail:* andrea.giuliani@inaf.it

We will describe the current status of the ASTRI Mini-Array, under construction at the Teide Astronomical Observatory in Tenerife, Spain. The final layout of the array will include nine small Cherenkov telescopes covering an area of about 650 x 270 square meters. The ASTRI telescopes adopt a dual-mirror Schwarzschild-Couder aplanatic optical design. In the focal plane, the ASTRI camera, based on silicon photo-multiplier detectors, will cover a large field-of-view ( $\geqslant$ 10 deg in diameter). This system also provides good gamma-ray sensitivity at very high energies (VHE, above 10 TeV) combined with a good angular resolution. The scientific goals of the ASTRI Mini-Array include spectral and morphological characterization of the LHAASO sources and other Pevatron candidates, studies of PWNe and TeV halos, Blazar monitoring at VHE, fundamental physics and follow-up of transient events. The beginning of the scientific operations is planned for mid 2025. The first three years will be dedicated to the core science and the ASTRI Mini-Array will be run as an experiment. It will gradually move towards an observatory model in the following years, open to the community.

[*]Speaker

# 1. The ASTRI project

The ASTRI Mini-Array is based on nine 4-meter class Cherenkov telescopes, currently under implementation at the Teide Observatory in Tenerife, Spain [1, 2] . The ASTRI collaboration is responsible for the array's realization and operation. It includes more than 150 researchers in 5 countries, encompassing several INAF institutes (in Bologna, Catania, Milan, Padua, Palermo, and Rome ), universities (Perugia, Padua, Catania in Italy, the University of Sao Paulo in Brazil, the North Western University in South Africa, the Geneva University in Switzerland), the Fundación Galileo Galilei and the Instituto de Astrofísica de Canarias. The array layout and altitude (2400 m) and the design of the telescopes are optimized for the observation of the gamma-ray sky in the 1-200 TeV energy band, extending the operational range of energies of the current imaging Cherenkov telescopes.

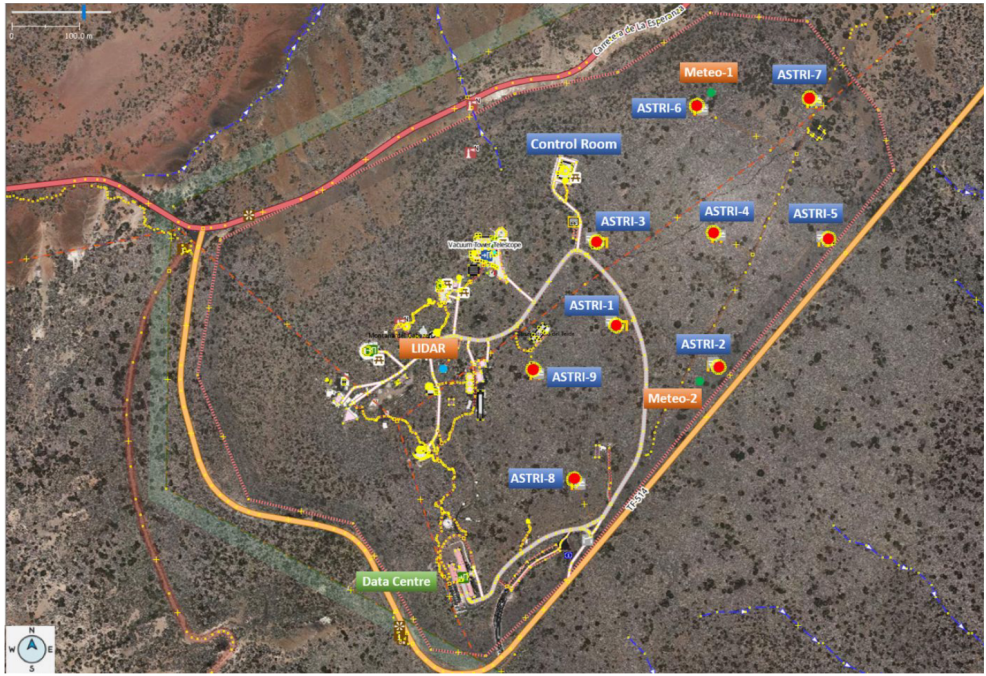

**Figure 1:** Future layout of the nine telescopes of the ASTRI Mini-Array at the Teide Observatory (IAC)

Since the start of the scientific operations, synergies with other gamma facilities in the northern hemisphere (LHAASO, HAWC, MAGIC, VERITAS, CTA-N ) are foreseen and encouraged.

A prototype of the ASTRI telescope, called ASTRI Horn[1], was installed on the slope of the Etna volcano in 2014. ASTRI Horn demonstrated the potentialities of the dual-mirror telescope and the ASTRI camera with a campaign of observations of the Crab Nebula (see [3]) the first gamma-ray source ever observed with a dual-mirror Cherenkov telescope.

---

[1]in honor of Guido Horn D'Arturo, the Italian astronomer who first proposed the use of tessellated mirrors for astronomy.

## 2. The ASTRI Mini-Array

The experience with the ASTRI Horn prototype was crucial for developing the Mini Array. The telescopes are an evolution of the telescope installed on Etna but with significant improvements. The electro-mechanical structure has been optimized for mass ( reduced by 30% ), functionality and maintainability. The coating of the mirrors was also improved using a combination of Al, $SiO_2$ and $ZrO_2$ (see figure 2). The optical design is based on a modified Schwarzschild-Couder configuration. ([4], see also [5]). This configuration allows us to obtain a better correction of aberrations at large incident angles, even for small focal ratios, and good angular resolution across the entire field of view. This optical system has the additional advantage of reducing the plate scale, allowing for smaller pixel sizes than usual Cherenkov telescopes and the consequent possibility to use solid-state photosensors such as the Silicon Photo-Multipliers (SiPMs).

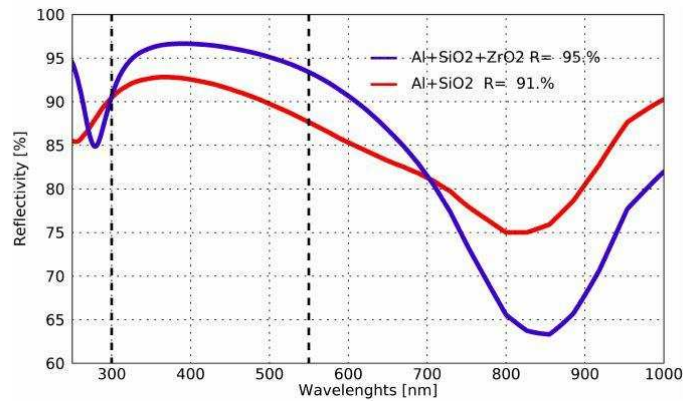

**Figure 2:** Comparison between the coating used for the mirrors of the ASTRI-Horn prototype (Al+$SiO_2$) and the one used for the ASTRI Mini-Array telescopes (Al+$SiO_2$ +$ZrO_2$). The vertical dashed lines mark the band where most of the Cherenkov light is emitted by the showers.

The camera system is also being improved evolving the prototype one mounted on ASTRI Horn, in the context of an industrial contract issued by INAF to the CAEN-EIE GROUP industrial pool. SiPM tiles of sensors are produced by Hamamatsu photonics. With a pixel size of 7x7 $mm^2$, they are grouped in matrices of 8x8 pixels; 37 matrices are then arranged to adapt to the curved focal plane of the telescope. The innovative electronics for peak detection developed by INAF for the camera (CITIROC ASICS, WEEROC) will ensure a small amount of data.

An interferential filter used as the front window (see [6] and [7]) reduces the sensitivity at wavelengths greater than 550 nm, where the contribution from the night sky background dominates over the Cherenkov light produced by the showers. The ASTRI telescopes work in stereoscopic mode relying on a layout covering a strip of about 650 x 270 square meters (see figure 1). The ASTRI Mini-Array software, developed by the team, is envisioned to handle an observing cycle in every aspect, namely observation preparation, observation execution, data processing and dissemination to the community. The ASTRI Mini-Array ICT (Information and Communication Technology) facilities will be located on both Tenerife Island and in Italy. Tenerife will include the Local Control Room, the On-Site Data Centre at the Teide Observatory and the Array Operation Center at IACTEC in La Laguna. Italy will host the Data Centre in Rome, plus a few Remote Array operation centres in other INAF institutes.

## 3. The ASTRI Mini-Array performance

The combination of telescope design, array layout and altitude of the site, makes the ASTRI Mini-Array best suited for the observation of the sky in the multi-TeV regime, with a good sensitivity up to 200 TeV. Figure 3 shows the ASTRI Mini-Array sensitivity for deep exposures (that will typically last hundreds of hours, see the following section). ASTRI Mini-Array will then detect fluxes of a few times $10^{-13}$ erg cm$^{-2}$ s$^{-1}$ for energies of a few tens of TeV. These fluxes are similar to those of the sources detected by particle shower detectors.

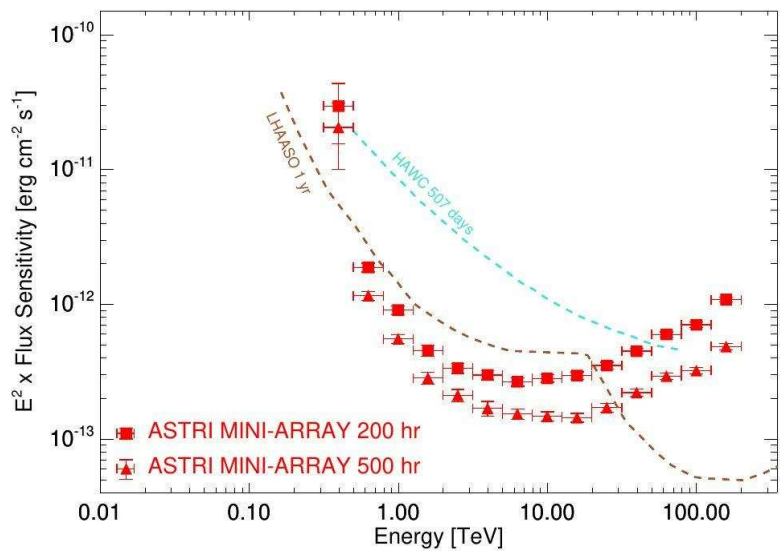

**Figure 3:** Sensitivity of the ASTRI Mini-Array for 200 and 500 hours of exposure (for a point-like source with a Crab-like spectrum). The HAWC and LHAASO sensitivity (for typical observation times) are also shown for comparison.

This sensitivity is combined with the good angular and energetic resolution typical of the Imaging Atmospheric Cherenkov Telescopes (IACTs). Montecarlo simulations [8] show indeed that the ASTRI Mini-Array can reach an angular resolution of 3 arcmin with an energy resolution of 10 - 15 %. These performances are required for better characterize the morphology of extended sources at the highest VHE (above 10 TeV ).

Another important feature of the ASTRI Mini-Array is its large field of view ($\geqslant$ 10 degrees in diameter) with an almost homogeneous off-axis acceptance. As shown in figure 4, the off-axis sensitivity is similar (within a factor 2) to the on-axis value for large angles. This feature will allow the ASTRI Mini-Array to study extended sources, perform surveys and, on Galactic fields, to collect exposure on more than one source at a time.

A set of ASTRI Mini-Array instrument response functions with format compatible with science tools packages like e.g. gammapy [9] and ctools [10], are already available on Zenodo.org[2].

---

[2]https://zenodo.org/record/6827882

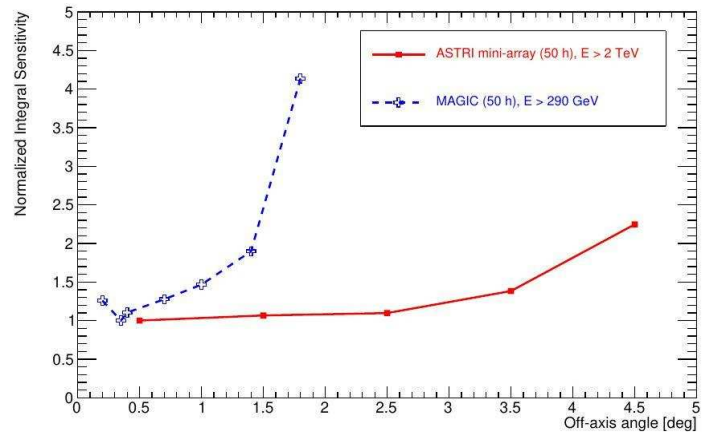

**Figure 4:** ASTRI Mini-Array sensitivity as a function of the off-axis angle for a point-like source (for energies greater than 2 TeV and 50 hours of exposure ). The MAGIC sensitivity is also shown for comparison with other IACTs

## 4. Scientific operations

During the first three years of operations the ASTRI Mini-Array will be run as an experiment. The ASTRI Science team will develop a strategy to concentrate the observational time on a limited number of programs with clearly identified objectives ( see [11] ). After this initial period the project will move towards an observatory model (see [12] and [13] ).

The scientific observations of the ASTRI Mini-Array will cover both Galactic and extragalactic targets. One of the main scientific goals will be the characterization and identification of the PeVatron candidates detected by particle shower detectors like HAWC or LHAASO [14]. Many of these sources are non-identified and possibly associated with many counterparts. In particular, ASTRI Mini-Array, thanks to its good angular resolution, will be able to investigate the morphology of these sources, to identify the astrophysical process at the basis of the emission, and then help to understand the nature of the Galactic PeVatrons. Other Galactic targets in the scientific program are Pulsar Wind Nebulae (PWN) and TeV Halos. The energy-dependent maps produced by the ASTRI Mini-Array observations will allow an understanding of how these sources work.

The ASTRI Mini-Array will contribute on the extragalactic side to monitor and characterise Blazar emission. The observation of these sources can also lead to the evidence of exotic processes of fundamental physics, like e.g. Lorentz Invariance Violation (LIV) or Axion-like particles (ALP) productions. ASTRI Mini-Array will also follow up transient events of particular interest (for example, events with known low redshift) like Gamma-Ray Bursts and Gravitational Events. The ASTRI Mini-Array science is not limited to $\gamma$-ray astrophysics, but it will include the measurement and study of primary Cosmic Rays.

All these studies require long exposures especially for energies larger than few tens of TeV. The observational strategy will then focus on few sky fields to obtain long exposures on selected targets. At this scope, the large field of view of the ASTRI Mini-Array will play an important role in allowing the observation (on the Galactic plane) of a few sources at a time. Observations at large zenith angles will also contribute to reaching the required statistics, especially in the multi-TeV energy range. Long deep exposures will be feasible also thanks to the characteristics of the ASTRI camera that allow observations in moderate moonlight conditions, that is, under a high night sky background rate. The effect of the moonlight on the sky brightness depends on various factors, mainly Moon altitude, phase and angular distance between the Moon and the target. This last aspect strongly depends on the target's sky position; for example, sources at high ecliptic latitudes tend always to have large distances from the Moon, reducing the impact of the moonlight also for large Moon-phases (first quarter and beyond). Observation with moonlight will increase the maximum observational time for a given target with respect to a pure dark time, up to 40 (60) per cent for low (high) ecliptic latitude sources.

## 5. Schedule

The Mini-Array is currently under construction at the IAC Teide observatory. The deployment of the telescopes foresees 4 phases:

- Phase 0: Installation of the first telescope structure (see figure 5), with final acceptance expected by February 2023

- Phase 1: Installation of the first three telescope structures (ASTRI 1, 8 and 9 in figure 1 ). Expected by June 2023

- Phase 2: Installation of the cameras on the first three telescopes. At the end of this phase (Late 2023), a preliminary observational campaign of gamma-ray sources will start.

- Phase 3: Installation of the remaining six telescopes. The entire array is expected to be ready for science verification by mid-2025 (albeit some delays are possible due to the current complicated international situation, which makes challenging the procurement of materials and electronic components)

The ASTRI Mini-Array scientific program will start with a first 3-year phase dedicated to the core science goals. In a second phase, the instrument will gradually open up to an observatory-type model, with guest observer programs open to the astronomical community in which a fraction of the time will be assigned to scientific proposals through a Time Allocation Committee procedure.

## 6. Conclusion

The ASTRI Mini-Array is under construction at the Teide observatory in Tenerife, Spain. Its implementation is expected in 2025 when an array of nine Cerenkov telescopes will be installed. The innovative design of the telescopes and the cameras will allow us to investigate with good angular and energetic resolution the most promising PeVatrons candidates and other Galactic and extragalactic sources up to energies of 200 TeV with an unprecedented field of view for a Cherenkov telescope array. The first three years of observations will be dedicated to a core science program (see [11]). More information can be found at the ASTRI website : http://astri.me.oa-brera.inaf.it/

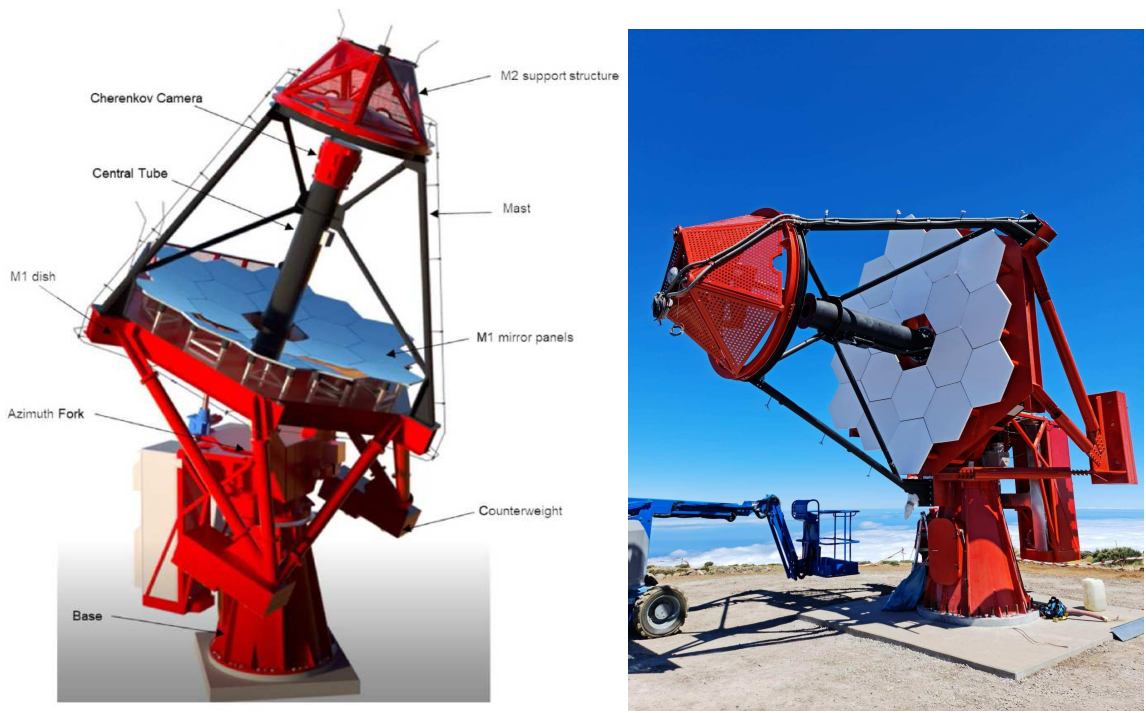

**Figure 5:** Left: A 3D model of an ASTRI telescope. Right: the ASTRI-1 telescope at the Teide site

**Acknowledgments**

This work is being carried out in the context of the ASTRI Project, mainly supported by INAF and the Italian Ministry for University and Research (MUR) We gratefully acknowledge support from the people, international agencies, and organizations listed here: http://www.astri.inaf.it/en/library/. This paper went through the internal ASTRI review process.

**References**

[1] G. Pareschi, *The ASTRI SST-2M prototype and mini-array for the Cherenkov Telescope Array (CTA)*, in *Ground-based and Airborne Telescopes VI*, H.J. Hall, R. Gilmozzi and H.K. Marshall, eds., vol. 9906 of *Society of Photo-Optical Instrumentation Engineers (SPIE) Conference Series*, p. 99065T, Aug., 2016, DOI.

[2] S. Scuderi, A. Giuliani, G. Pareschi, G. Tosti, O. Catalano, E. Amato et al., *The ASTRI Mini-Array of Cherenkov telescopes at the Observatorio del Teide*, *Journal of High Energy Astrophysics* **35** (2022) 52.

[3] S. Lombardi, O. Catalano, S. Scuderi, L.A. Antonelli, G. Pareschi, E. Antolini et al., *First detection of the Crab Nebula at TeV energies with a Cherenkov telescope in a dual-mirror Schwarzschild-Couder configuration: the ASTRI-Horn telescope*, **634** (2020) A22 [1909.12149].

[4] V. Vassiliev, S. Fegan and P. Brousseau, *Wide field aplanatic two-mirror telescopes for ground-based γ-ray astronomy*, Astroparticle Physics **28** (2007) 10 [astro-ph/0612718].

[5] G. Sironi, *Aplanatic telescopes based on Schwarzschild optical configuration: from grazing incidence Wolter-like x-ray optics to Cherenkov two-mirror normal incidence telescopes*, in *Society of Photo-Optical Instrumentation Engineers (SPIE) Conference Series*, S.L. O'Dell and G. Pareschi, eds., vol. 10399 of *Society of Photo-Optical Instrumentation Engineers (SPIE) Conference Series*, p. 1039903, Sept., 2017, DOI.

[6] G. Romeo, G. Bonanno, G. Sironi and M.C. Timpanaro, *Novel silicon photomultipliers suitable for dual-mirror small-sized telescopes of the Cherenkov telescope array*, Nuclear Instruments and Methods in Physics Research A **908** (2018) 117 [1806.00703].

[7] O. Catalano, M. Capalbi, C. Gargano, S. Giarrusso, D. Impiombato, G. La Rosa et al., *The ASTRI camera for the Cherenkov Telescope Array*, in *Ground-based and Airborne Instrumentation for Astronomy VII*, C.J. Evans, L. Simard and H. Takami, eds., vol. 10702 of *Society of Photo-Optical Instrumentation Engineers (SPIE) Conference Series*, p. 1070237, July, 2018, DOI.

[8] S. Lombardi, L.A. Antonelli, C. Bigongiari, M. Cardillo, S. Gallozzi, J.G. Green et al., *Performance of the ASTRI Mini-Array at the Observatorio del Teide*, in *37th International Cosmic Ray Conference*, p. 884, Mar., 2022, DOI.

[9] C. Deil, R. Zanin, J. Lefaucheur, C. Boisson, B. Khelifi, R. Terrier et al., *Gammapy - A prototype for the CTA science tools*, in *35th International Cosmic Ray Conference (ICRC2017)*, vol. 301 of *International Cosmic Ray Conference*, p. 766, Jan., 2017 [1709.01751].

[10] J. Knödlseder, M. Mayer, C. Deil, J.B. Cayrou, E. Owen, N. Kelley-Hoskins et al., *GammaLib and ctools. A software framework for the analysis of astronomical gamma-ray data*, **593** (2016) A1 [1606.00393].

[11] S. Vercellone, C. Bigongiari, A. Burtovoi, M. Cardillo, O. Catalano, A. Franceschini et al., *ASTRI Mini-Array core science at the Observatorio del Teide*, Journal of High Energy Astrophysics **35** (2022) 1.

[12] A. D'Aì, E. Amato, A. Burtovoi, A.A. Compagnino, M. Fiori, A. Giuliani et al., *Galactic observatory science with the ASTRI Mini-Array at the Observatorio del Teide*, Journal of High Energy Astrophysics **35** (2022) 139.

[13] F.G. Saturni, C.H.E. Arcaro, B. Balmaverde, J. Becerra González, A. Caccianiga, M. Capalbi et al., *Extragalactic observatory science with the ASTRI mini-array at the Observatorio del Teide*, Journal of High Energy Astrophysics **35** (2022) 91.

[14] Z. Cao, F.A. Aharonian, Q. An, L.X. Axikegu, Bai, Y.X. Bai, Y.W. Bao et al., *Ultrahigh-energy photons up to 1.4 petaelectronvolts from 12 γ-ray Galactic sources*, **594** (2021) 33.